\documentclass[useAMS,usenatbib]{mn2e}
\usepackage{graphicx}
\usepackage{epstopdf}
\usepackage{subfigure}

\newcommand{\bvec}[1]{\mbox{\boldmath ${#1}$}}
\newcommand{\vdot}[2]{\left<{#1}\,|{#2}\right>}
\newcommand{\abs}[1]{\mbox{$|{#1}|$}}

\newcommand{\var}[1]{{\rm Var}\left[{#1}\right]}
\newcommand{\cov}[2]{{\rm Cov}\left[{#1},{#2}\right]}

\newcommand{\fracd}[2]{\mbox{$\frac{\displaystyle{#1}}{\displaystyle{#2}}$}}

\newcommand{\Dx}{\Delta x}
\newcommand{\Dy}{\Delta y}
\newcommand{\DF}{\Delta F}
\newcommand{\Dxj}{\Delta x_j}
\newcommand{\Dyj}{\Delta y_j}
\newcommand{\DFj}{\Delta F_j}

\newcommand{\Dj}{\bvec{D}_j}
\newcommand{\Pj}{\bvec{P}_j}

\newcommand{\R}{\bvec{R}}
\newcommand{\D}{\bvec{D}}
\newcommand{\PSF}{\bvec{P}}
\newcommand{\PZ}{\bvec{P}_0}
\newcommand{\Px}{\bvec{P}_{x}}
\newcommand{\Py}{\bvec{P}_{y}}

\title[Difference imaging photometry]{Difference imaging photometry of blended gravitational microlensing events with a numerical kernel}
\author[M.D. Albrow et al.]{M.D. Albrow$^{1}$\thanks{E-mail:Michael.Albrow@canterbury.ac.nz (MDA)},
K. Horne$^{2}$, D.M. Bramich$^{3}$, P. Fouqu\'e$^{4}$, V.R. Miller$^{1}$, 
\newauthor
J.-P. Beaulieu$^{5}$, C. Coutures$^{6,5}$, J. Menzies$^{7}$, A. Williams$^{8}$, V. Batista$^{5}$, D.P. Bennett$^{9}$, 
\newauthor
S. Brillant$^{10}$, A. Cassan$^{11}$, S. Dieters$^{5}$, D. Dominis Prester$^{12}$,  J. Donatowicz$^{13}$, 
\newauthor
J. Greenhill$^{14}$, N. Kains$^{2}$, S. R. Kane$^{15}$, D. Kubas$^{10}$, J.B. Marquette$^{5}$,  
K.R. Pollard$^{1}$, 
\newauthor
K. C. Sahu$^{16}$, Y. Tsapras$^{17,18}$, J. Wambsganss$^{11}$, M. Zub$^{11}$\\ \\
$^{1}$Beatrice Tinsley Institute\thanks{http://www.beatricetinsleyinstitute.org/}, Department of Physics and Astronomy, University of Canterbury, Private Bag 4800, Christchurch, \\ New Zealand \\
$^{2}$SUPA, School of Physics and Astronomy, University of St. Andrews, North Haugh, St Andrews, KY16 9SS, United Kingdom \\
$^{3}$Isaac Newton Group of Telescopes, Apartado de Correos 321, E-38700 Santa Cruz de la Palma, Canary Islands, Spain\\
$^{4}$LATT, Universit«e de Toulouse, CNRS, 14 avenue Edouard Belin, F-31400 Toulouse, France\\
$^{5}$Institut dÕAstrophysique de Paris, CNRS - UMR 7095, 
Universite Pierre \& Marie Curie, 98bis Bd Arago, 75014 Paris, France\\
$^{6}$CEA, DSM, DAPNIA, Centre d'\'Etudes de Saclay, 91191 Gif-sur-Yvette Cedex, France \\
$^{7}$South African Astronomical Observatory, PO Box 9, 7935 Observatory, South Africa \\
$^{8}$Perth Observatory, Walnut Road, Bickley, Perth 6076, Australia \\
$^{9}$University of Notre Dame, Department of Physics, 225 Nieuwland Science Hall, 
Notre Dame, IN 46556, USA  \\
$^{10}$European Southern Observatory, Casilla 19001, Vitacura 19, Santiago, Chile\\
$^{11}$Astronomisches Rechen-Institut (ARI), Zentrum f¬ur Astronomie (ZAH), Heidelberg University, 
M¬onchhofstrasse 12-14, \\ 69120 Heidelberg, Germany \\
$^{12}$Department of Physics, University of Rijeka, Omladinska 14, 51000 Rijeka,
Croatia \\
$^{13}$Technical University of Vienna, Dept. of Computing, Wiedner Hauptstrasse 10, Vienna, Austria\\
$^{14}$University of Tasmania, School of Maths and Physics, Private bag 37, GPO Hobart, Tasmania 7001, Australia\\
$^{15}$NASA Exoplanet Science Institute, Caltech, MS 100-22, 770 South Wilson 
Avenue Pasadena, CA 91125, USA\\
$^{16}$Space Telescope Science Institute, 3700 San Martin Drive, Baltimore, MD 21218, USA\\
$^{17}$Las Cumbres Observatory Global Telescope Network, 6740 Cortona Drive, 
Suite 102, Goleta, CA 93117, USA \\
$^{18}$Astrophysics Research Institute, Liverpool John Moores University, 
Liverpool CH41 1LD, UK 
}

\date{Released 2008 April}

\begin{document}

\date{Accepted       Received ; in original form }

\pagerange{\pageref{firstpage}--\pageref{lastpage}} \pubyear{2008}

\maketitle

\label{firstpage}

\begin{abstract}
The numerical kernel approach to difference imaging has been implemented and applied to gravitational microlensing events observed by the PLANET collaboration. The effect of an error in the source-star coordinates is explored and a new algorithm is presented for determining the precise coordinates of the microlens in blended events, essential for accurate photometry of difference images.
It is shown how the photometric reference flux need not be measured directly from the reference image 
but can be obtained from measurements of the difference
images combined with knowledge of the statistical flux uncertainties. The improved performance of the new algorithm, relative to ISIS2, is demonstrated.
\end{abstract}


\section{Introduction}

Over the last 15 years, gravitational microlensing \citep{Einstein1936} has been observed routinely and used in the study of dark baryonic matter \citep{Alcock1993, Aubourg1993} and stellar atmospheres \citep{Albrow1999, Albrow2001a, Albrow2001b, Fields2003, Cassan2004},  and in the search for extrasolar planets \citep{Albrow2000, Albrow2001c, Gaudi2002, Bond2004, Udalski2005, Beaulieu2006, Dong2008, Gaudi2008}.
The PLANET collaboration \citep{Albrow1998} operates a number of 1-m class telescopes distributed around the Southern Hemisphere and performs round-the-clock CCD photometry of microlensing events that have been discovered and alerted in real time by the OGLE \citep{Udalski1994, Udalski2003} and MOA \citep{Bond2002} microlensing surveys.

In this paper we discuss recent advances in the PLANET difference imaging reduction pipeline, focussing on 
several subtleties inherent in the reduction of blended microlensing events. In what follows we use the convention
\begin{equation}
D = \frac{R \otimes K - T}{\sum_{ij} K_{ij}}
\,
\label{Eqn1}
\end{equation}
for difference image $D$, reference image $R$, target image $T$ and convolution kernel $K$. That is, 
 a difference image is defined as the convolved reference minus the target, normalised so that it is on the effective exposure scale of the reference.

We illustrate the methods using a sample dataset of images of microlensing event OGLE 2008-BLG-229 that were taken using the Elizabeth 1.0-m telescope at the South African Astronomical Observatory during PLANET operations in 2008. The microlensing event was alerted by OGLE on 2008 May 3 and initially was predicted to have low magnification. 
Subsequent observations revealed  a blended moderate magnification event, peaking with magnification $A_{0} = 7.24$ on 2008 July 18. OGLE data and parameters for the event can be obtained from the OGLE Early Warning System website \footnote{http://ogle.astrouw.edu.pl/ogle3/ews/ews.html}. 
The SAAO observations consist of 84 images, spanning the time from 7 days before maximum until 23 days after maximum.

\section{The reduction pipeline}

\subsection{Introduction}

In our first years of operation, PLANET photometry was performed both in real-time at the telescopes and offline using the DoPHOT PSF-fitting code \citep{Schechter1993} under a reduction pipeline 
written mainly by JPB. Following the development of the ISIS code \citep{Alard1998}, we adopted the 
difference imaging method for obtaining our best photometry offline, while still employing the DoPHOT pipeline 
at the telescope sites. In the 2006 season, we began using a difference imaging photometric pipeline at the telescope sites and for our final offline photometry. The offline version, known as pySIS2, was developed by MDA, and is based on the ISIS 2 code of \citet{Alard2000}. An adaptation of pySIS by CC, known as WISIS, is used for most of the real-time at-telescope reductions, while the pySIS2 code is used at the Perth Observatory. The pipelines allow single images to be reduced immediately after observation using an existing reference template. As better quality images are acquired, the reference template can be updated and the previously observed images rereduced.

 \subsection{Image registration}
 
The ISIS code requires that all images be fully registered to an astrometric reference. Bright stars are
located on all frames and cross-correlation of their positions followed by an iterative rejection scheme is
used to define an astrometric transformation for each target image.

An innovation introduced to our offline pipeline in 2007 was the removal of the requirement to fully register images.
Instead, we register the images only by a shift in X and Y to the nearest pixel, thus avoiding the need for interpolation and resampling.
Resampling is generally undesirable since it introduces correlations between adjacent pixels, meaning that their flux uncertainties are no longer described by Poisson statistics. In the case of images that are close to or below critical spatial sampling, resampling introduces an artifact where stellar PSF's are not constant or slowly-varying across an image, but depend on the subpixel location of their centroids. Such images usually do not 
subtract cleanly.
Integer-pixel registration was handled in 
our modified version of ISIS by offsetting
the kernel centroid by the subpixel registration residual. We note that this approach is somewhat less flexible than the standard ISIS code, in that it cannot work with sets of images with 
rotations relative to each other. 

\subsection{Difference imaging with a numerical kernel}

In 2008, we have developed a new version of the code, pySIS3, that is no longer based on ISIS image subtraction. Instead, for the difference-imaging step, we have implemented the algorithm of \citet{Bramich2008}. 
In this method, the kernel is represented as a numerical pixel array, rather than the decomposition of
Gaussians multiplied by  polynomials used in ISIS. The numerical kernel is able to accommodate images
with irregular PSFs, for instance trailed images, that ISIS cannot cope with. An implicit feature of the method is
that complete registration is not required and the kernel naturally incorporates subpixel offsets. Image registration in pySIS3 is hence restricted to integer pixel shifts.
\citet{Bramich2008} shows examples of how the new algorithm outperforms ISIS. DMB's code has
been used successfully to discover new variable stars in the globular cluster NGC 6366 \citep{Arellano2008}.

Our implementation has been used for the analysis of several microlensing events, appearing in forthcoming papers on MOA 2007-BLG-197 \citep{Cassan2009}, OGLE 2004-BLG-482 \citep{Zub2009}, and OGLE 2007-BLG-472 \citep{Kains2008}, and for a transit search \citep{Miller2009}. There are several specific details of our implementation that we note here. 

First, the algorithm is generally more computationally intensive than ISIS, and the computation time scales strongly with the number of pixels required 
for the kernel array. For microlensing events, we generally reduce only a subsection of
the images, typically 250 x 250 pixels centred on the microlens. Our plate scales are typically around
0.3 arcsec/pixel.

Second, results depend on the size of the pixel array chosen for the kernel. The kernel needs to be large enough to encapsulate the transformation between reference and target, but not so large that it introduces noise
into the convolution. We have found the best results by employing a circular kernel with a radius (in pixels) given by
\begin{equation}
R_{\rm kernel} = \min \left( 7, 4 \left( {\rm FWHM_{target} - FWHM_{reference}  } \right) \right)
\,
\end{equation}
where ${\rm FWHM_{target}}$ and  ${\rm FWHM_{reference}}$ are the full-width at half-maxima (in pixel units)  of the microlens on the target and reference images.
Regions of the kernel that are located more that 7 pixels from its centre are represented by 3x3 binned pixels in order to reduce noise. The values for the binned kernel pixels are computed from the equations in \citet{Bramich2008} but using a $3 \times 3$ boxcar-smoothed version of the reference image.

Third, to prevent saturated stars that have irregular PSFs from entering the kernel determination, we mask a
circular area of radius 15 pixels around all pixels that are saturated in either the reference or target image as well
as masking the microlens itself. The default behavior is to use all the remaining image pixels to determine the kernel using the \citet{Bramich2008} algorithm. In cases where images are contaminated by artifacts, such as diffraction spikes, that are not easily masked, we have found that using `stamps' around bright unsaturated stars rather than the entire unmasked image renders a kernel that is less prone to contamination.

Fourth, the photometric scaling factor,
\begin{equation}
s \equiv \sum_{ij} K_{ij}
\,
\label{Eqn2}
\end{equation}
where $K$ is the convolution kernel, represents the relative difference in {\it effective} exposure time between
the reference and target images, i.e. it accounts for differences in both exposure time and atmospheric transparency. For each target, if $s$ is significantly different from the ratio of 
true exposure times, this usually indicates a poorly subtracted target or one affected by cloud.

Fifth, for images that have poor spatial sampling - either close to or even below critical sampling - we
use the following technique. The registered images are oversampled  by a factor of two in each direction
using cubic O-MOMS interpolation \citep{Blu2001}. This type of resampling does not transfer flux across original-pixel boundaries. 
A stack of typically 10 of the best of these images are then mapped onto the best-seeing image and combined to make a reference image. Since our individual images usually have random subpixel dithers, this process generally results in an oversampled reference so long as the initial undersampling is not too severe. 
This approach is similar to that employed in 
R. Gilliland's code for difference-imaging of undersampled {\it HST} WFPC2 images \citep{Gilliland2000, 
Albrow2001d}. We note that this approach is still under development and is not used for the sample data set
of images for OGLE 2008-BLG-229 in this paper, which are not undersampled.

\subsection{Photometry}

To extract photometric measurements from our difference images, we first use the Bphot program from ISIS
to compute the PSF of the reference image. This PSF is then convolved with the previously-computed kernel to produce a PSF for each target image. The PSF is then normalised and resampled at the subpixel lens coordinates
using cubic O-MOMS interpolation. Any residual background is removed from the difference image using a
low-order polynomial model.
Finally, the PSF is fitted to the difference image using optimal extraction, i.e  each pixel weighted by the inverse of its flux variance.

\subsubsection{Flux errors from imprecise coordinates}

\begin{figure}
\includegraphics[scale=0.45]{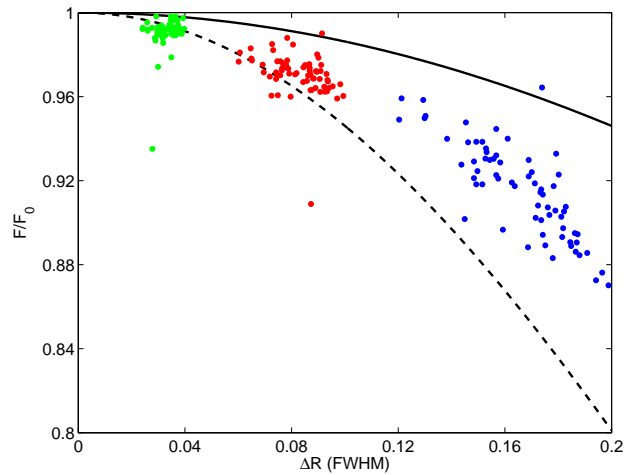}
\caption{Flux determined at offset coordinates relative to the flux at the correct coordinates as a function of coordinate offset, $\Delta R$ in units of PSF full-width at half-maximum (FWHM). Solid line for unweighted PSF-fitting, dashed line for optimal PSF-fitting with zero background, both under the assumption of a Gaussian PSF. The three distinct groups of data points are computed from the SAAO difference images for OGLE 2008-BLG-229, with the PSF shifted in $x$ by 0.2 (green), 0.5 (red)  and 1.0 pixels (blue) (0.06, 0.16, 0.31 arcsec respectively) from their correct value.}
\label{fluxerr}
\end{figure}

\begin{figure}
\includegraphics[scale=0.4]{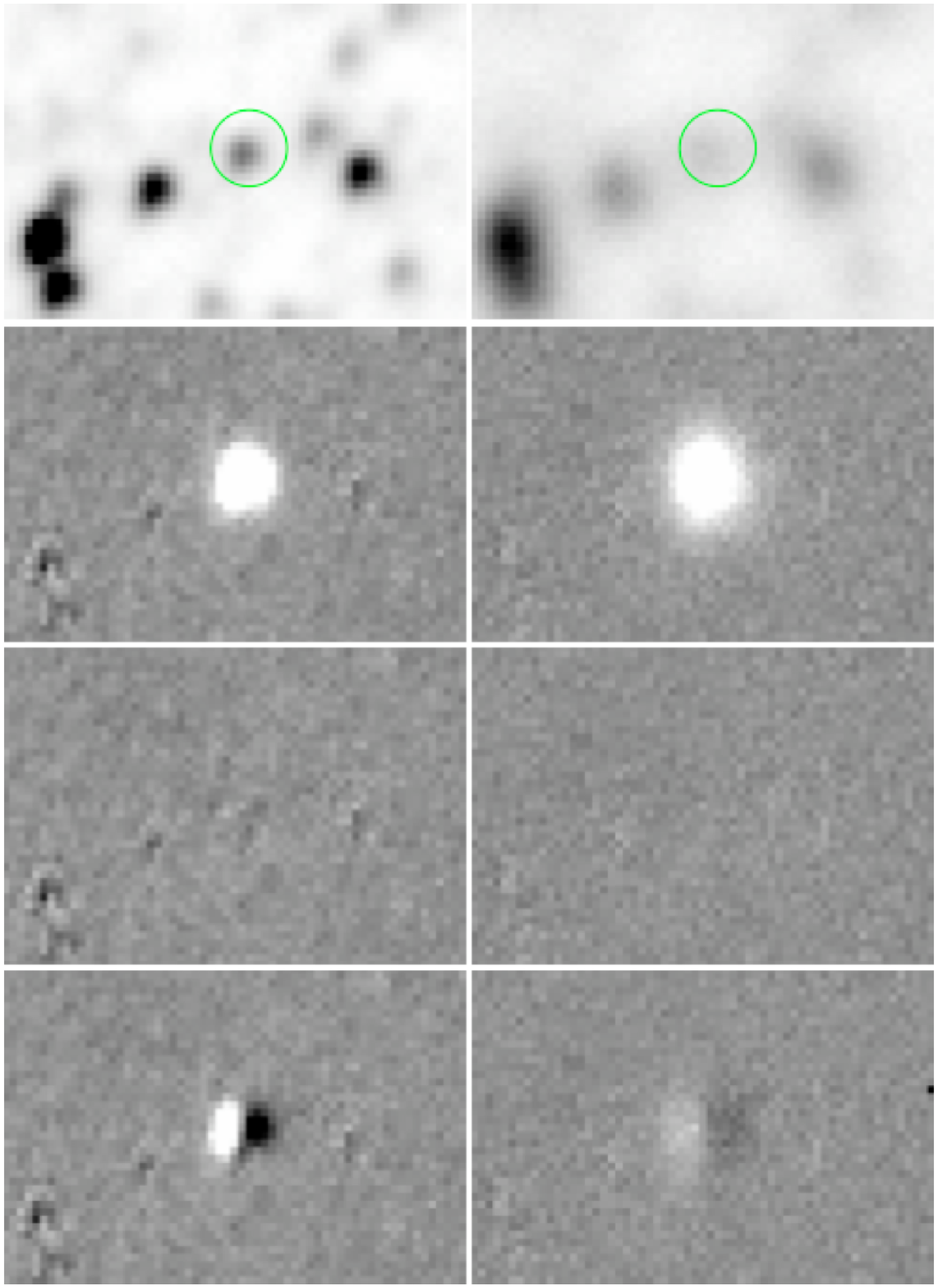}
\caption{Sample images 78 (left; good seeing) and 83 (right; poor seeing)  from the SAAO observations of OGLE 2008-BLG-229.
Top row: direct images registered to the nearest pixel, linear greyscale encompassing 95\% of pixel values.
Second row: difference images, linear greyscale range -400 to +400.
Third row: difference images after subtraction of PSF fitted at correct coordinates.
Bottom row: difference images after subtraction of PSF fitted at coordinates shifted by +1 pixel in X.}
\label{images}
\end{figure}

All of the microlensing events towards the Galactic Bulge are blended to some extent in regular ground-based imaging, i.e. the PSF of the microlensed source star overlaps with nearby stars. 
One implication of this is that the coordinates of the true source star are often displaced by perhaps several tenths of an arcsec from the centroid of the PSF.

When performing photometry on a set of difference images by PSF-fitting, small errors in target location
lead to systematic underestimates of the flux. In Figure~\ref{fluxerr} we show how the flux measurement depends
on coordinate displacement, $\Delta R$ (measured in units of FWHM), under a Gaussian PSF assumption.
The two limiting cases, shown as lines, are (a) optimal PSF fitting in the
zero-background limit and (b) unweighted PSF fitting, essentially the background-limited case. 
In both limits, the flux error scales as the square of the ratio of the coordinate error to the FWHM of the PSF.
This means that a flux error due to an incorrectly-positioned PSF is
more serious for images with better seeing and that such a coordinate error introduces scatter into a lightcurve derived from a set of variable-seeing (or even variable-background) images.

We can demonstrate this effect using our test case of SAAO OGLE 2008-BLG-229. We have created difference images using as a reference template a single good-seeing image, number 18, acquired close to the peak of the microlensing event.
We have performed optimal photometry on the set of difference images at the correct coordinates (i.e. zero-offset) and again with the coordinates shifted in X by 0.2, 0.5 and 1.0 pixels (0.06, 0.16, 0.31 arcsec) from their correct position. 
The three displayed groups of data points in Figure~\ref{fluxerr} show the flux measured at the offset
coordinates relative to the flux at the correct position for the three sets of displaced-coordinate measurements. 
The figure shows that, as predicted, the measured difference flux decreases and its dispersion increases with coordinate offset.

It is important to note that, for difference-image photometry, the above effect applies to difference fluxes, $\Delta F_{i}$. The total flux is given by $F_{i} = F_{0} - \Delta F_{i}$, where $F_{0}$ is the flux on the reference image, $R$ in Eqn (1). This means that the place in the lightcurve where coordinate errors are manifested most strongly depends on the choice of reference image. The flux error is largest for regions of the lightcurve where the magnification is most different from that of the reference image. Consequently, such errors can be minimized for a given part of a lightcurve 
(for instance some part of a lightcurve suspected to display an anomaly) by choosing a reference image where the source star has a similar magnification. 

\subsubsection{Precise coordinates for blended events}

For high-magnification events, the true source location may be discerned from images taken near peak magnification, when the flux from the true microlensed source star dominates that from nearby blended stars. For data sets comprised of images that have precise registration, a sum of the absolute values
of the difference images can be used successfully to refine the coordinates from their initial estimate, even for relatively low-magnification events. This method was used in pySIS2.

In our current circumstance, we have sets of images that are registered only to the nearest pixel. The method of stacking
difference images could, in principle, be applied to this data, provided each of the difference images is first
shifted by the subpixel registration residual. However, such a shift requires accurate knowledge of the subpixel
residuals and involves interpolation and resampling, 
a process that is inaccurate for images with near- or below-critical spatial sampling.

A better way, that retains the original sampling, is to use the residuals from PSF-fits to the difference images.
In appendix~\ref{A1} we introduce a new algorithm to refine the source-star coordinates by minimising
these residuals over all images. In our photometric code, the algorithm operates as an integral part of the measurement process.

In Figure~\ref{images} we show direct, difference and residual images for  two sample observations, numbered 78 (good seeing) and 83 (poor seeing) in the SAAO data set for OGLE 2008-BLG-229. Both images were 
taken during the final days of data, when the source was at a magnification, $A \approx 2.5$. The reference
image was again image number 18, taken near peak magnification, $A \approx 7$.

Our coordinate algorithm resulted in a change of 0.54 pixels (0.17 arcsec) in the location of the target star relative to the position found from our best-seeing image (which was adopted as the astrometric reference). 
Using the correct coordinates, the residual images (difference images after subtraction of the fitted PSF) 
are very clean.  

The effect of a coordinate offset of 1 pixel (0.3 arcsec)  is to produce a residual image with large positive and negative flux features remaining. The effect of such an offset on the lightcurve can be tested.  We compare difference-flux lightcurves in a blend-free way by mapping them to a point-source point-mass lens model as follows. At time $t_{i}$,
the unblended flux from the source star is given by
\begin{equation}
F_{i} = A_{i} F_{\rm base} = F_{0} - \Delta F_{i}
\,
\end{equation}
where $A_{i}$ is the magnification that we constrain to be defined by the OGLE geometric parameters for the event ($u_{0}$ = 0.139, $t_{\rm E}$ = 53.994~d, $t_{0}$ = JD2454665.780), $F_{0}$ is the unblended source flux on the reference image and $F_{\rm base}$ is the unblended baseline source flux. For each lightcurve, we solve for $F_{0}$ and $F_{\rm base}$by minimizing
\begin{equation}
\chi^{2} \equiv \sum_{i} \frac{ \left( A_{i} F_{\rm base}  - F_{0} + \Delta F_{i} \right)^{2} }{\sigma_{i}^{2}}
\end{equation}

\begin{figure}
\includegraphics[scale=0.45]{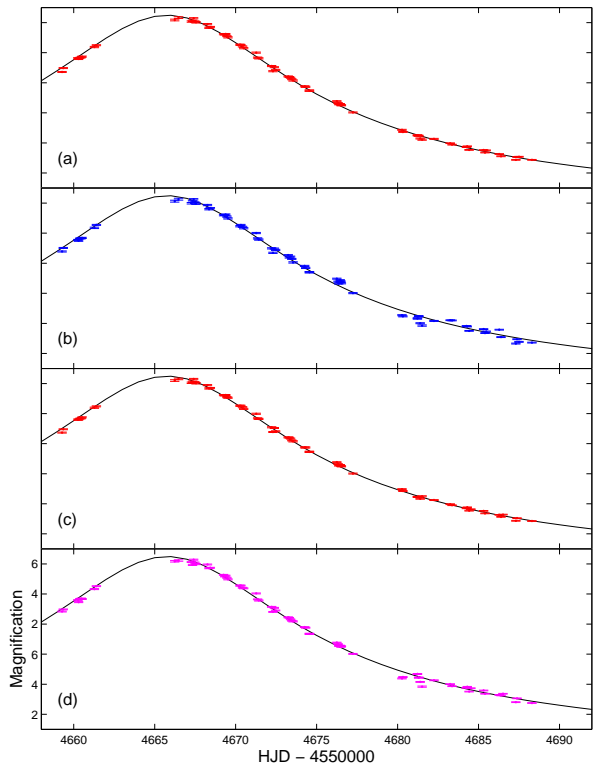}
\caption{Lightcurves for OGLE 2008-BLG-229: (a) using single template image from near peak,
(b) as for (a) but with a 1 pixel coordinate offset, (c) combination of 10 images for reference template, 
(d)  best lightcurve obtained using ISIS2. }
\label{lightcurves}
\end{figure}

\begin{figure}
\includegraphics[scale=0.45]{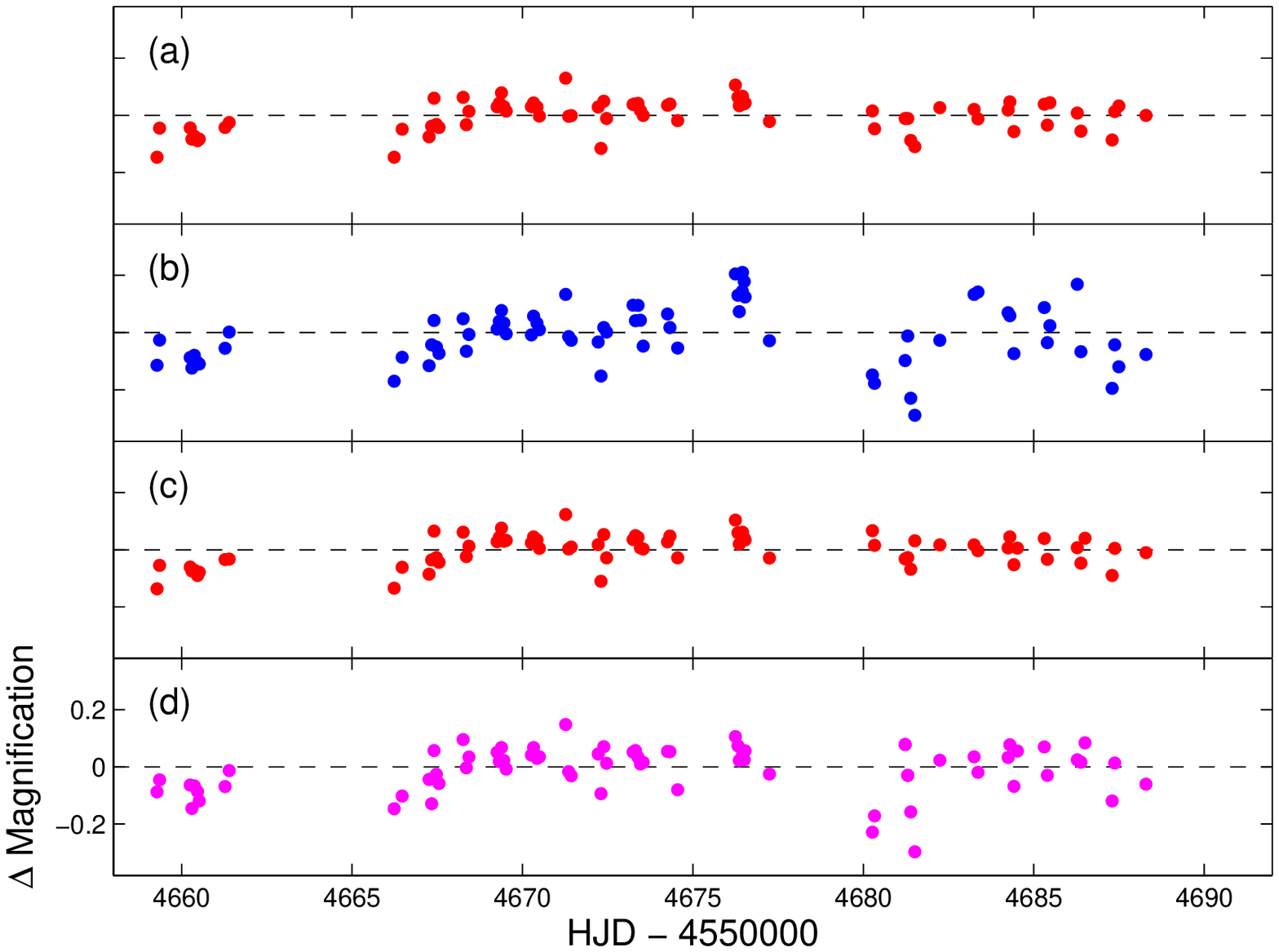}
\caption{Lightcurve residuals for OGLE 2008-BLG-229 corresponding to Fig~\ref{lightcurves} after
subtraction of a best-fit PSPL model with the geometric parameters set to be those found from 
OGLE. Errorbars are suppressed for clarity.}
\label{lcresiduals}
\end{figure}

Lightcurves for the whole data set are shown in Figure~\ref{lightcurves} for the correct coordinates and for those with a 1-pixel offset. An increased scatter is visible in
the lightcurve corresponding to the offset coordinates. In the same figure, we also show the best lightcurves
we have derived using a template created from the 10 best-seeing images using pySIS3 and ISIS2. Figure~\ref{lcresiduals} shows the corresponding residuals.
The effect of a coordinate error can be seen through comparison of the upper two panels, particularly during the last 10 days data points when the magnification is most different from that of the reference image.
The superior performance of the new numerical-kernel
algorithm can be discerned through comparison of the two lower panels, where the ISIS2 lightcurve residuals 
(panel (d)) have a 53\% greater RMS scatter than the pySIS3 residuals (panel (c)). We note that there is
an apparent systematic residual in all the displayed lightcurves, where the earliest data points lie below the 
magnification curve. This is likely due to the fact that we have constrained, rather than fitted, the 
underlying geometric model.

\subsubsection{Reference flux}

The output of our difference-image photometry is the difference in flux, $\Delta F_{i}$, between each 
target image, $i$, and a photometric reference image. The reference image may be a single observation, as in the preceding sections, but more commonly 
is created from a combination of images with the best seeing and lowest sky background. In order to interpret
our observations we require the flux, $F_{i} = F_{0} - \Delta F_{i}$, where $F_{0}$ is the flux on the reference image. In a model-dependent sense, a deblended $F_{0}$ can be derived as a fitting parameter as done above.
More usually, particularly during the time when observations are being acquired, $F_{0}$ is measured from a PSF-fit at the lens coordinates on the reference image. Often this 
estimation is in error due to the crowded nature of the Galactic Bulge fields in which we observe.

An incorrect value for $F_{0}$ can, to some extent, be compensated for in the microlens blend fraction,
\begin{equation}
f_{\rm bl} \equiv \frac{F_{s}}{F_{s} + F_{b}}
\,
\end{equation}
where for magnification $A$, the baseline flux $(F_{s} + F_{b})$
increases to 
\begin{equation}
F = F_{s} A +F_{b} =  \left( F_{s} + F_{b} \right) \left( f_{\rm bl} A + \left( 1 - f_{\rm bl} \right) \right)
\ 
\end{equation}
and the blend fraction is derived from lightcurve fitting. 
However, blending parameters so derived are inconsistent between different datasets for the same event, 
may take on non-physical values, 
and certainly no-longer have the correct physical interpretation as the fraction of light contributing to the stellar
PSF at baseline from the microlensing source star. 

A more-successful approach that we have developed is to  choose $F_{0}$ so that the photometric uncertainties 
in the set of $\Delta F_{i}$ measurements are consistent with Poisson noise for fluxes $F_{i} = F_{0} - \Delta F_{i}$.
The algorithm for this determination, referred to as the {\it Poisson reference flux} is detailed in Appendix~\ref{A2}. 
The method uses information from all suitable images and the variance in the Poisson reference flux scales 
roughly with the inverse of the number of images. This variance is generally smaller than the variance for a direct flux measurement from the reference image, which scales with the inverse of the number of individual images incorporated into the reference.
For our sample lightcurve, the reference flux is  measured directly as $F_{0} = 101,820 \pm 90$ ADU (likely to include extra blended light), while the poisson method yields $F_{0} = 93,570 \pm 40$ ADU.

\section{Summary}

Difference imaging has proved to be a powerful technique in the measurement of gravitational microlensing
flux variations and for variable stars and transiting extrasolar planets. The numerical kernel method
introduced by \citet{Bramich2008} represents  a significant advance over the analytic kernel of \citet{Alard1998}.

In this paper we have shown that, for blended microlensing events, precise determination of the coordinates of the microlens is necessary to obtain accurate photometry. We have presented a new algorithm, based on photometric residuals, to measure such coordinates to high precision. 

Additionally,  we have introduced a new method to measure the reference flux in a manner that does not depend on an (often inaccurate) analysis of the photometric reference image. The {\it Poisson reference flux} method 
produces a more accurate and precise determination of the unmagnified source flux than a direct measurement from the reference image.

A comparison has been made between photometry from our new code and photometry using ISIS for SAAO images of microlensing event OGLE 2008-BLG-229. The new code produces measurements that display significantly less scatter about a point-source point-mass-lens lightcurve based on the OGLE-determined geometric parameters for the event.
 
\section*{Acknowledgements}

This work was supported by the Marsden Fund of New Zealand under contract UOC302 and
by ANR grant HOLMES during MDA's visit to the  Observatoire Midi-Pyr\'en\'ees, Toulouse, in
May - June 2008. We thank the referee, Scott Gaudi, for his thorough report.

\appendix

\section{Refining the lens position}
\label{A1}

\subsection{ target position from a single difference image }

We wish to find sub-pixel offsets, $\Dx$ and $\Dy$, that
create the best match of a PSF model $\PSF$ to a difference image $\D$.
Expand the PSF model to first order in $\Dx$ and $\Dy$,
\begin{equation}
	\PSF = \PZ + \Dx\, \Px + \Dy\, \Py
\ ,
\end{equation}
and thereby write the
residual image\footnote{Note that $\Dx>0$ shifts the PSF peak to smaller $x$,
and similarly for $y$. We adopt this sign convention to simplify the equations.}
as
\begin{equation}
	\R = \D - \DF \, \PSF
\ ,
\end{equation}
where $\PZ$ is the unshifted PSF, $\DF$ is the difference flux, and
the $x$ and $y$ gradients of the unshifted PSF are
\begin{equation}
	\Px \equiv \fracd{ \partial \PZ }{ \partial x }
\ , \hspace{5mm}
	\Py \equiv \fracd{ \partial \PZ }{ \partial y }
\ .
\end{equation}
The PSF is normalised to
\begin{equation}
	\sum_i P_i = 1
\ 
\end{equation}
when summed over pixels $i$.
With $\sigma^2_i= \var{D_i}$ the variance of the data $D_i$ in pixel $i$,
the $\chi^2$ statistic measuring the ``badness-of-fit'' is
\begin{equation}
	\chi^2 = \sum_i \left( \fracd{R_{i}}{\sigma_{i}} \right)^2
	= \abs{\R}^2
\ .
\end{equation}
Here we adopt the convenient notation
\begin{equation}
	\vdot{\bvec{A}}{\bvec{B}}
	\equiv \sum_i \fracd{A_i\,B_i}{\sigma^2_{i}}
\hspace{5mm} \abs{A}^2 \equiv \vdot{\bvec{A}}{\bvec{A}}
\end{equation}
for the inverse-variance weighted ``dot product'' of ``image vectors''
$\bvec{A}$ and $\bvec{B}$, and note
that $\chi^2$ is the squared norm of the residual image $\R$.

Starting with $\DF=\Dx=\Dy=0$,
we calculate the difference flux $\DF$, for fixed $\Dx$ and $\Dy$,
by 
optimally scaling the shifted PSF $\PSF$ to fit the difference image $\D$,
giving
\begin{equation}
	\DF = \fracd{ \vdot{\D}{\PSF} }{ \abs{\PSF}^2 }
\ ,
\end{equation}
and the corresponding variance
\begin{equation}
	\var{\DF} = \fracd{ 1 }{ \abs{\PSF}^2 }
\ .
\end{equation}
Next we update the image position, for fixed $\DF$, by scaling the PSF gradient 
images to fit the residuals,
\begin{equation}
\Dx \rightarrow \Dx + \fracd{ \vdot{ \R}{\Px} }{ \DF\, \abs{\Px}^2 }
\end{equation}
\begin{equation}
\Dy \rightarrow \Dy + \fracd{ \vdot{ \R}{\Py} }{ \DF\, \abs{\Py}^2 }
\end{equation}
with variances
\begin{equation}
\var{\Dx} = \fracd{ 1 }{ \left(\DF\right)^2\, \abs{\Px}^2 }
\end{equation}
\begin{equation}
\var{\Dy} = \fracd{ 1 }{ \left(\DF\right)^2\, \abs{\Py}^2 }
\end{equation}

The above results minimise $\chi^2=\abs{\R}^2$
if $\DF$, $\Dx$ and $\Dy$ are independent, and if $\sigma_i$ are fixed.
As these assumptions are only approximately true, iteration is required.
We find that 
the iteration is faster and more stable if we take account of
$\R$ being a linear function of $\Dx$ and $\Dy$. We then have
two coupled equations,
\begin{equation}
\Dx = \fracd{ \vdot{ \D - \DF\,\left( \PZ + \Dy\, \Py \right)} {\Px} }
	{ \DF\, \abs{\Px}^2 }
\ ,
\end{equation}
\begin{equation}
\Dy = \fracd{ \vdot{ \D - \DF\,\left( \PZ - \Dx\, \Px \right)} {\Py} }
	{ \DF\, \abs{\Py}^2 }
\ .
\end{equation}
Write these in matrix form as
\begin{equation}
H \cdot
	\left( \begin{array}{c} \DF\,\Dx \\ \DF\,\Dy \end{array} \right)
	= 
	\left( \begin{array}{c} \vdot{\D-\DF\,\PZ}{\Px}
	\\ \vdot{\D-\DF\,\PZ}{\Py} \end{array} \right)
\ ,
\end{equation}
with the Hessian matrix
\begin{equation}
H = \left( \begin{array}{cc} \abs{\Px}^2 & \vdot{\Px}{\Py}
	\\ \vdot{\Px}{\Py} & \abs{\Py}^2 \end{array} \right)
\ .
\end{equation}
The solution is 
\begin{equation}
	\left( \begin{array}{c} \DF\, \Dx \\ \DF\, \Dy \end{array} \right)
	= 
H^{-1} \cdot
	\left( \begin{array}{c} \vdot{\D-\DF\,\PZ}{\Px}
	\\ \vdot{\D-\DF\,\PZ}{\Py} \end{array} \right)
\ .
\end{equation}
where the inverse of the Hessian matrix is
\begin{equation}
H^{-1} = \fracd{1}{\det(H)}
\left( \begin{array}{cc}
	\abs{\Py}^2 & -\vdot{\Px}{\Py}
	\\ -\vdot{\Px}{\Py}  & \abs{\Px}^2
\end{array} \right)
\ ,
\end{equation}
with the Hessian determinant
\begin{equation}
\det(H) = \abs{\Px}^2\, \abs{\Py}^2 - \vdot{\Px}{\Py}^2
\ .
\end{equation}
The sub-pixel shift is then
\begin{equation}
\label{eqn:dx}
\Dx = \fracd{
	\vdot{\D-\DF\,\PZ}
	{ \left( \abs{\Py}^2 \Px - \vdot{\Px}{\Py}\, \Py \right) }
	}{
	\DF\, \det(H)
	}
\ ,
\end{equation}
\begin{equation}
\Dy = \fracd{
	\vdot{\D-\DF\,\PZ}
	{ \left( \abs{\Px}^2 \Py - \vdot{\Px}{\Py}\, \Px \right) }
	}{
	\DF\, \det(H)
	}
\ .
\end{equation}

Since $H^{-1}$ is the parameter covariance matrix,
the diagonal elements give the variances
\begin{equation}
\label{eqn:vardx}
\var{\Dx} = \fracd{ \abs{\Py}^2}{ \left( \DF \right)^2\, \det(H) }
\ ,
\end{equation}
\begin{equation}
\var{\Dy} = \fracd{ \abs{\Px}^2 }{ \left( \DF \right)^2\, \det(H) }
\ ,
\end{equation}
and the off-diagonal element gives the covariance
\begin{equation}
\cov{\Dx}{\Dy} = \fracd{ - \vdot{\Px}{\Py} }
	{ \left( \DF \right)^2\, \det( H ) }
\ .
\end{equation}

Note that with $\DF$ in the denominator, these expressions become
problematic when $\DF\approx0$. Such images carry very little information
about the target location. Fortunately, when we
optimally average over several images, the inverse-variance weights
shift the $\DF$ factors to the numerator,
so that these images receive low weight.

\subsection{ lens position from many images }

In fitting a microlensing dataset, we have many difference images $\Dj$,
and the corresponding PSFs $\Pj$.
The above analysis provides estimates (with error bars)
that we correspondingly 
label $\DFj$ for the difference fluxes,
$\Dxj$ and $\Dyj$ for the sub-pixel offsets.

The difference fluxes are different for each image, but the sub-pixel
shift establishing the lens position on the reference image
should be the same for all images.
The optimal average of the estimates $\Dxj$ from individual images is
\begin{equation}
\Dx = \fracd{\sum_j w_j\, \Dxj }
	{ \sum_j w_j }
\hspace{5mm}
\var{\Dx} = \fracd{1 }
	{ \sum_j w_j }
\end{equation}
with inverse-variance weights $w_j = 1 / \var{\Dxj}$.
Explicit evaluation using (\ref{eqn:dx}) and (\ref{eqn:vardx}) gives
\begin{equation}
\Dx = \fracd{
	\sum_j \DF\, \vdot{\D-\DF\,\PZ}
		{ \Px - \fracd{\vdot{\Px}{\Py}}{\abs{\Py}^2}\,\Py }
	}{
	\sum_j \left(\DF\right)^2 \left( \abs{\Px}^2 - 
		\fracd{\vdot{\Px}{\Py}^2}{\abs{\Py}^2} \right)
	}
\ ,
\end{equation}
\begin{equation}
\var{\Dx} = \fracd{ 1
	}{
	\sum_j \left(\DF\right)^2 \left( \abs{\Px}^2 - 
	\fracd{\vdot{\Px}{\Py}^{2}}{\abs{\Py}^2} \right)
	}
\ .
\end{equation}
The corresponding expressions for $\Dy$ and $\var{\Dy}$
are found by reversing $x$ and $y$.
For clarity we omit the index $j$ that labels
every term in the sums over images $j$.

Note that the $\DF$ factors appear in the numerator only,
so that difference images with $\DFj\approx 0$ are included
in the sums but with appropriately low weight. In our implementation, the 
algorithm typically converges to $\sim 10^{-3}$ pixels in $\sim 4$ iterations.

\section{Computing the poisson reference flux}
\label{A2}

We retain here the convention (Eqn.~\ref{Eqn1}) where a difference image is on the same effective exposure scale as the reference image and a negative difference flux results when the target star is brighter than it is on the reference. The expected pixel-integrated star flux on a target image is then
\begin{equation}
\left< F \right> = \left( F_{0} - \Delta F \right) s
\,
\end{equation}
where $F_{0}$ is the pixel-integrated flux of the lens star in the reference image (ADU), $\Delta F$ is the pixel-integrated differential flux of the lens star in the difference image (ADU) and $s$ is the exposure scale factor between the reference image and the target image (Eqn.~\ref{Eqn2}).
Assuming a noiseless reference image, the variance in flux of the lens on a single difference image, is given approximately by
\begin{equation}
\var{F} = \frac{N_{\rm pix}  \, \sigma_{0}^{2}}{g^{2} \, s^{2}} + \frac{N_{\rm pix} \, F_{\rm sky}}{g \, s^{2}} + \frac{F_{0} -\Delta F}{g \, s} 
\ 
\label{NoiseEqn}
\end{equation}
where $\sigma_{0}^{2}$ is the readout noise variance $(e^{-}/{\rm pix})^{2}$, $g$ is the gain ($e^{-}/$ADU), 
$F_{\rm sky}$ is the background flux (ADU/pixel) on the target image and $N_{\rm pix}$ is the effective number of pixels in the photometric aperture.
The different denominators for each of the three terms in Eqn.~\ref{NoiseEqn} are due to the read-out-noise being measured in units of electrons, $F_{\rm sky}$ in units of target-frame ADU and $F_{0} - \Delta F$ in units of reference-frame ADU. 
An estimate of the reference flux from this single difference image is therefore
\begin{equation}
F_{0} = g \, s \, \var{F} - \frac{N_{\rm pix} \, \sigma_{0}^{2}}{g \, s}  - \frac{N_{\rm pix} \, F_{\rm sky}}{s} + \Delta F
\ 
\end{equation}
with variance
\begin{equation}
\var{F_{0}} \simeq \var{\Delta F} 
\ .
\end{equation}
An optimal estimate for $F_{0}$ is obtained by combining such measurements from all target images,
\begin{equation}
<F_{0} > = \frac{ \sum_{j} \frac{F_{0,j}}{\var{F_{0,j}}} } {\sum_{j} \frac{1}{\var{F_{0,j}}} } 
\ ,
\end{equation}
with associated variance,
\begin{equation}
\var{<F_{0} >} = \frac{ 1 } {\sum_{j} \frac{1}{\var{F_{0,j}}} } 
\ .
\end{equation}
For aperture photometry measurements, $N_{\rm pix}$ is the number of pixels in the photometric aperture. In the case of optimal PSF-fitting photometry, $N_{\rm pix}$ is the effective number of sky pixels, which we define to be equal to the
variance in the flux measurement that is due to the background divided by the background variance per pixel.
To estimate $N_{\rm pix}$, consider the background-limited case, where the noise variance is the same on each pixel, $V_0$, and is the dominant contributor to the variance in the flux measurement.  If our optimal extraction is confined to some aperture, then the variance in the measured flux is
\begin{equation}
\var{\Delta F}  = \frac{ \left( \sum_{x,y} P(x,y) \right)^{2} } 
{  \sum_{x,y} \frac{P(x,y)^{2}}{V_{0}} } \equiv N_{\rm pix} V_{0}
\ ,
\end{equation}
where $P(x,y)$ is the PSF and hence
\begin{equation}
N_{\rm pix} = \frac{ \left( \sum_{x,y} P(x,y) \right)^{2} } 
{  \sum_{x,y} P(x,y)^{2} }
\ .
\end{equation}
Under a Gaussian PSF,
\begin{equation}
P(r) = \frac{1}{2 \pi \sigma^{2}} e^{-r^{2}/2 \sigma^{2}}
\ ,
\end{equation}
where $\Delta$ is the full-width at half-maximum,
\begin{equation}
N_{\rm pix} = \frac{\left( \int P\,dx\,dy\right)^{2}}{\int P^{2}\,dx\,dy}
	    = 4\,\pi\,\sigma^{2} = \frac{\pi\,\Delta^{2}}{2\ln{2}}
\ .
\end{equation}
equivalent to an aperture of radius $2\sigma$. For a Gaussian truncated
at $r=R$, this equates to
\begin{equation}
N_{\rm pix} = 4 \pi \sigma^{2} \frac{ (1 - e^{-R^{2}/2 \sigma^{2} })^{2}} { 1 - e^{-R^{2}/\sigma^{2}} } 
    = \frac{ \pi \Delta^{2}}{2 \ln 2} \frac{ (1 - e^{-4 \ln 2 R^{2}/\Delta^{2} })^{2}} { 1 - e^{-8 \ln 2 R^{2}/\Delta^{2}} } 
\ .
\end{equation}

\end{document}